\documentclass[twocolumn,footinbib,prb,preprintnumbers,amsmath,amssymb]{revtex4}

\usepackage{graphicx}
\usepackage{graphics}
\usepackage{dcolumn}
\usepackage{bm}

\begin{document}

\title{Twist glass transition in regioregulated poly(3-alkylthiophene)s}

\author{Koji Yazawa}
\affiliation{Department of Biomolecular Engineering, 
Tokyo Institute of Technology. 
4259 Nagatsuta-cho, Midori-ku, Yokohama, Kanagawa 226-8501, Japan}

\author{Takakazu Yamamoto}
\affiliation{Chemical Resources Laboratory, 
Tokyo Institute of Technology. 
4259 Nagatsuta-cho, Midori-ku, Yokohama, Kanagawa 226-8503, Japan}

\author{Yoshio Inoue}
\affiliation{Department of Biomolecular Engineering, 
Tokyo Institute of Technology. 
4259 Nagatsuta-cho, Midori-ku, Yokohama, Kanagawa 226-8501, Japan}

\author{Naoki Asakawa}%
\thanks{Corresponding Author}
\email{nasakawa@bio.titech.ac.jp}
\affiliation{Department of Biomolecular Engineering, 
Tokyo Institute of Technology. 
4259 B-55 Nagatsuta-cho, Midori-ku, Yokohama, Kanagawa 226-8501, Japan}%

\date{\today}

\begin{abstract}
The molecular structure and dynamics of regioregulated poly(3-butylthiophene) (P3BT),
poly(3-hexylthiophene)(P3HT), and poly(3-dodecylthiophene) (P3DDT) were investigated
using Fourier transform infrared absorption (FTIR),
 solid state $^{13}$C nuclear magnetic resonance (NMR),
 and differential scanning calorimetry (DSC) measurements.
In the DSC measurements, the endothermic peak was obtained around 340 K in P3BT,
and assigned to enthalpy relaxation that originated from the glass transition
 of the thiophene ring twist in crystalline phase
from results of FTIR,
$^{13}$C cross-polarization and magic-angle spinning (CPMAS) NMR,
$^{13}$C spin-lattice relaxation time measurements, 
and centerband-only detection of exchange (CODEX) measurements.
We defined this transition as {\it twist-glass transition},
which is analogous to the plastic crystal - glassy crystal transition.

\end{abstract}

\maketitle


\section{Introduction}
Poly(3-alkylthiophenes) (P3ATs) 
show great promise as electronic and 
optoelectronic devices such as light-emitting diodes and field-effect transistors.
Particularly, the regioregular versions of P3ATs have received much attention 
because they show high crystallinity and high conductivity because of improvement of 
main chain packing. 
 
Their structures have been studied extensively to gain understanding of the physical properties of P3ATs. In particular, polymorphic behavior among three solid phases has been a main topic in structural investigation
\cite {Prosa1996,Meille1997,Park1997,Bolognesi2001,Causin2005}:
 Phase I, mainly taking an end-to-end arrangement for side chains;
Phase II, molecular packing with interdigitation, in which the
side chains are intercalated;
 and the nematic mesophase.
The realization of the phases depends strongly on the alkyl side chain length and
their thermal histories. Although it is said that the P3ATs with longer alkyl chains 
are likely to form Phase II, correlation between the side chain length and
the resultant phase has never been clearly determined. 

The dynamic structure of P3AT is important to find out physical properties 
such as thermochromism, in which the twist motion of the thiophene ring is thermally excited \cite{Inganas}.
{\it Conformon} is an elemental excitation related to such molecular structural defects
that it behaves as an interrupter of $\pi$-conjugation.
Therefore, the dynamics as well as dimension and density of conformons are related closely to 
the conjugation length of the $\pi$ electron \cite{Corish1997,Iwasaki1994}.
No studies are available for the molecular dynamics 
of P3ATs in spite of its importance.
 
Phases with conformon are called quasi-ordered phases \cite{Cheng Yang1996}. 
Typical examples of the quasi-ordered phase include liquid crystals (LCs)
 and plastic crystals (PCs).
Generically, LCs are found in rod-type molecules having translational freedom, whereas PCs are 
realized in spherical molecules such as C$_{60}$, having
translational symmetry, but rotational freedom.
Particular attention has been devoted to the glass transitions of LCs
 \cite{Tsuji1971,Johari1982,Chen1991,Ahumada1996,delCampo2002,Chen2003,Tokita2004}
 and PCs \cite{Adachi1968,Tycko1994,Pietrass1997,Amoureux1998,Decressain2005}
to obtain information about
the dynamics of quasi-ordered phases. 
Interestingly, some quasi-ordered phases
 can be frozen into a state called
 glassy liquid crystal (GLC) and glassy crystal (GC) if LCs and PCs are cooled rapidly enough. 
Investigating these phases helps us determine
the dynamics of translational or rotational orders.
They can serve as model systems
for dynamics of the glass transition of amorphous polymer. 
Dynamics of quasi-ordered phases in polymers, however, have not been investigated
and little is known on whether the glassy states produced from the quasi-ordered phases
exist or not. 

In this study, we shall give attention to
the dynamics of quasi-ordered phases, particularly twist motion of thiophene main chain
in the crystalline state of P3ATs.
At first, we show an endothermic peak for poly(3-butylthiophene)(P3BT) in DSC measurements,
and peak shifts of C$_\beta$-H out-of-plane deformation region in FTIR measurement around 340 K.
Next, we also investigate the dynamics of P3BT around the transition
by $^{13}$C CPMAS NMR ,$^{13}$C spin-lattice relaxation time measurements ($T_{1 \rm C}$),
and centerband-only detection of exchange(CODEX) measurements,
and show evidence of the glass transition with respect to the twist motion.
We define this transition as {\it twist glass transition}, which is an example analogous to
PC-GC transition.


\section{Experiments}


\subsection{Materials}

Regioregular HT (head-to-tail)-type poly(3-butylthiophene) (P3BT),
 poly(3-hexylthiophene)(P3HT), and poly(3-dodecylthiophene) (P3DDT) were purchased 
from Rieke Metals Inc. \cite{Rieke} and were used without further purification. 

\subsection{DSC Measurements}

The DSC thermograms of the powder P3ATs were recorded mainly on a differential scanning calorimeter (Pyris Diamond; PerkinElmer Inc.).
 Rapid quenching using liquid nitrogen and the second heating scan were
performed on another calorimeter (Seiko DSC 220; Seiko Corp.). 
The respective masses of all samples were 3--6 mg.
The temperature and 
heat flow scales were carefully calibrated at different heating rates
using an indium standard with nitrogen purging.

\subsection{FTIR measurements}

The FTIR measurements were carried out using an FTIR microscope (AIM-8800; Shimadzu Corp.)
 that was equipped with an FTIR hotstage (LK-600; Linkam Scientific Instruments, Ltd.) and a cooling unit (L600A). All spectra were recorded at a resolution of 2 cm$^{-1}$ and with an accumulation of 16 scans.
Samples cast on BaF$_2$ substrates from the chloroform solution and dried in an oven under vacuum for 12 h were heated and cooled at a constant rate of 
10 K/min using nitrogen purging.

\subsection{Solid-state NMR Measurements}

We carried out variable temperature $^{13}$C cross-polarization and magic-angle spinning (CPMAS) \cite{Stejskal},
$T_{1{\rm C}}$, and centerband-only detection of exchange (CODEX) \cite{deAzevedo1999}
 measurements using a NMR spectrometer (GSX-270; JEOL) operating
 at 270 MHz for $^1$H and 67.9 MHz for $^{13}$C.
For CPMAS measurements, experimental conditions were set up with
90$^{\circ}$ pulse length of 5.2 $\mu$s, contact time of 2 ms,
recycle delay of 10 s, and the MAS rate of 3.2 kHz.
The $^{13}$C chemical shifts were referenced externally to the methyl carbon resonance of hexamethylbenzene at 17.36 ppm.
We used Torchia method \cite{Torchia} for
$T_{1{\rm C}}$ measurements.
For CODEX measurements, experimental conditions were set up with
90$^{\circ}$ pulse length of 5.0 $\mu$s, contact time of 2 ms,
recycle delay of 10 s, and the MAS rate was monitored by fiber optics and set at 5000$\pm$2Hz.
Pure-exchange CODEX spectra are obtained by measuring a reference spectrum
with dulations of $t_m$ and $t_z$ interachanged and subtracting the CODEX spectrum
from it. 
Original powder samples were used.
We also carried out variable temperature proton transverse relaxation time
 ($T_{2\rm H}$) measurements.
The detail experimental conditions are in Supplementary Information.
\subsection{Chemical shielding calculation}

The $^{13}$C chemical shielding tensors of a model compound for
HT poly(3-alkylthiophene) 
were calculated using an
{\it ab initio} self-consistent field (SCF) coupled Hartree-Fock method
with gauge invariant atomic orbitals (SCF-GIAO)\cite{giao,giao2}.
The HT-trimeric(3-methylthiophene)
was employed for all calculations. Structural optimization assuming s-trans conformation and SCF-GIAO shielding calculations were carried out using
the 6-31G(d) basis set.
All {\it ab initio} chemical shielding calculations were
performed using a Gaussian 98 (Rev.A7) program package\cite{g98}
run on a Cray C916/12256 supercomputer
at the Computer Center, Tokyo Institute of Technology.


\section{Results and Discussion}


\subsection{DSC Measurements}

\begin{figure}[htbp]
\begin{center}
\includegraphics[scale=0.4]{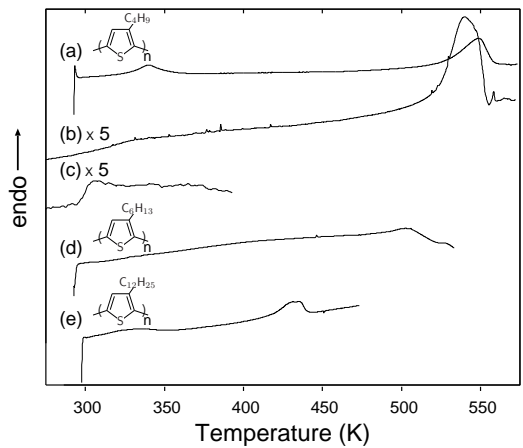}
\caption{DSC charts for powder samples of the first heating scan for P3BT (a), P3HT (d), and P3DDT (e),
and the second heating scan for P3BT (b), and (c).  The heating rate was 10 K/min.
After the first heating scan, the P3BT were cooled at {\it ca.} 50 K/min (b) or quenched by liquid nitrogen (c) 
and the second scan was run.}
\label{fig1}
\end{center}
\end{figure}

Figure \ref{fig1} shows DSC thermograms of the P3ATs measured on heating
at a rate of 10 K/min. 
Endothermic peaks were visible near 545 K, 500 K, and 435 K for P3BT, P3HT and P3DDT, respectively.
These transitions are attributed to the melting of crystalline
to the isotropic phase \cite {Cheng Yang1996}.
For P3DDT, another endothermic peak was observed around 330 K,
 which is consistent with the previous DSC study \cite {Cheng Yang1996}.
It is attributed to the melting of ordered side chains. 
Interestingly, P3BT gave rise to another endothermic transition around
 340 K.
 It is difficult to assign this transition to the melting of side chains
 because the butyl group is inferred to be in a liquid-like state 
at that temperature because P3HT shows no similar peak over the range between 223 K and 500 K. 
Causin \textit{et al.} reported similar DSC results and 
deduced from the results of wide angle x-ray diffraction (WAXD)
of P3BT that the transition is a polymorphic behavior from
the coexisting state between Phase II (interdigitation packing)
and Phase I (end-to-end packing), to Phase I
\cite {Causin2005}.
In the following, we will show evidence that the 340 K peak originates
from a twist glass transition. 
Their study showed that the traces of diffraction 
conditions exhibited drastic changes between 323 K and 373 K.
At 373 K, the (100) diffraction, attributed to side-by-side alignment, 
became sharper and more intense. The (010) signal 
(d$_{010}$ = 3.8\AA\  ), attributed to thiophene ring stacking, 
emerged from a very broad halo.
On the other hand, the WAXD patterns of Phase II were not clear at 323 K; 
i.e. the (100) diffraction was
weak and both Phase I and Phase II of (010) signals were not visible.
 Therefore, we inferred that no clear evidence exists to indicate polymorphism
between Phase I and Phase II.

Chen \textit{et al.} observed a similar endothermic peak
 at 332 K for regiorandom P3BT
and attributed it to the glass transition temperature \cite{Cheng S.-A1992}.
However, they did not identify its type of glass transition.
Figure \ref{fig1}(b) and Fig. \ref{fig1}(c) respectively show the second heating scans that were performed after the first heating scans
and two versions of quenching procedures, with cooling at 50 K/min and 
quenching by liquid nitrogen. 
We present a clear heat capacity jump that is believed to be a glass transition 
at 300 K in Fig. \ref{fig1}(c).
 Figure \ref{fig1}(b) shows no such jump,
 indicating that sufficiently rapid cooling prevents crystallization
 and gives rise to the conventional glass transition
by cage effect\cite{Letz2000}.
Therefore, we conclude that the both endothermic peaks at 332K
which is observed by Chen et al. and at 340K by us are
distinguishable from the conventional glass transition.
Slight difference between two temperatures of Chen's and ours, 332K and 340K, 
are probably because of the difference of regioregularity.
Since the regioregularity of our sample is
higher than Chen's, the transition temperature would be
higher due to the larger size of lamellar stacking.

\subsection{Temperature Dependence of FTIR Spectra}

\begin{figure}[htbp]
\begin{center}
\includegraphics[scale=0.3]{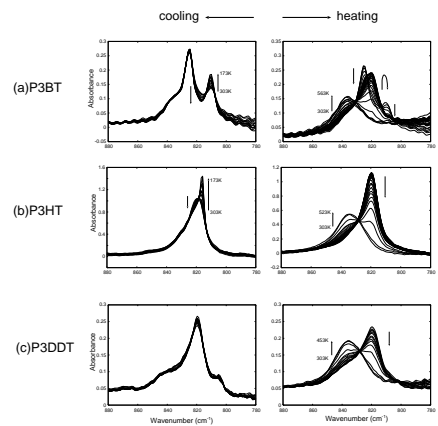}
\caption{Temperature dependence of infrared absorption spectra of thin films for P3BT (a), P3HT (b), and P3DDT (c) thin films. Only the C$_\beta$-H out-of-plane deformation band region is shown.}
\label{fig2}
\end{center}
\end{figure}

We carried out {\it in situ} FTIR
measurements on a thin film of P3BT to clarify the transition of P3BT around 340 K. Figure \ref{fig2} shows FTIR absorption spectra of 
P3BT, P3HT, and P3DDT in the
C{$_{\beta}$}-H out-of-plane deformation region as a function of temperature. 
First, we specifically examine the behavior at temperatures greater than 303 K (right side of Fig. \ref{fig2}). 
For P3HT and P3DDT, single peaks were observed at 820 cm$^{-1}$ at 303 K.
A new band 
grew at 836 cm$^{-1}$, whereas the 820 cm$^{-1}$ band
 decreased when both polymers were heated over the melting temperature of long range order of the main chain (Fig. \ref{fig2}(b) and Fig. \ref{fig2}(c)).
This phenomenon was consistent with
the previous FTIR study for regiorandom P3HT \cite {Zerbi}. 
We can respectively assign two bands, 820 cm$^{-1}$ and 836 cm$^{-1}$, to some sort of crystal and isotropic liquid.
 
\begin{figure}[htbp]
\begin{center}
\includegraphics[scale=0.40]{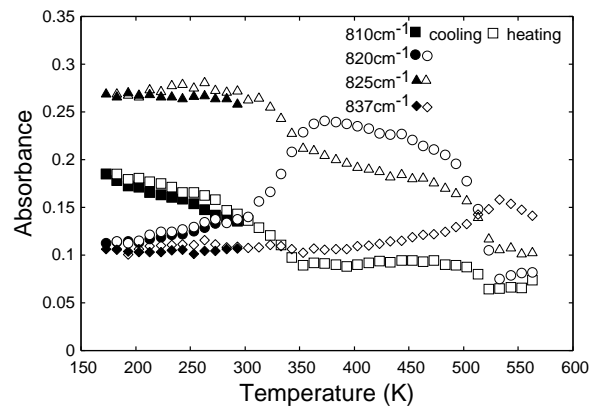}
\caption{Change in the absorption of the C$_\beta$-H out-of-plane deformation band for P3BT.
At first, the sample was cooled from 303 K to 173 K, and after 3 min storage heated from 173 K to 573 K with the rate of 10 K/min.}
\label{fig3}
\end{center}
\end{figure}

Unlike those, P3BT showed somewhat complex spectral changes.
Figure \ref{fig3} shows temperature dependent absorbance 
at each wavenumber for P3BT.
 At 303 K, the strongest peak 
 is at 825 cm$^{-1}$ and a small shoulder is observed at 810 cm$^{-1}$.
 With heating, the intensity of
the 820 cm$^{-1}$ band grows while both 825 cm$^{-1}$ and 810 cm$^{-1}$ bands decrease. 
It is noteworthy that the temperature range at which spectral changes occur in the FTIR spectra 
are consistent with results of DSC measurement.
From 353 K to 473 K, the band maximum remained at 820 cm$^{-1}$.
Around melting temperature (540 K), the 820 cm$^{-1}$ band disappeared, 
whereas the 836 cm$^{-1}$ band grew, which resembles the melting behavior of P3HT and P3DDT.
Therefore, the transition around 340 K must be a solid-solid transition because the melting temperature is much higher than 340 K. It differs, however, from a conventional glass transition
 mentioned by Cheng \cite {Cheng S.-A1992} 
 because, below the transition temperature, P3BT shows an x-ray diffraction peak 
 at 7$^{\circ}$, which corresponds to the (100) plane 
because of the lamellar structure \cite{Causin2005}. 
As mentioned above for the result of DSC, the second heating scan shows the heat capacity jump at 300 K, which is distinct from
 340 K (Fig. \ref{fig1}(c)).  

We also carried out FTIR measurements below room temperature (303 K $\sim$ 173 K) to elucidate the bands at 825 cm$^{-1}$ and 810 cm$^{-1}$.
The left side of Fig. \ref{fig2} shows the temperature dependence of FTIR spectra 
for P3BT, P3HT, and P3DDT with cooling.
Changes in the absorption of each peak (825 cm$^{-1}$, 820m$^{-1}$, and 810 cm$^{-1}$)
for P3BT are shown in Fig. \ref{fig3}.
In the lower temperature region, the absorbance at 825 cm$^{-1}$ was approximately constant.
 However, the band at 820 cm$^{-1}$ showed a gradual decrease.
 In addition, a small shoulder peak was apparent at 810 cm$^{-1}$.
Changes between the two bands showed a clear isosbestic point: the component attributed to
820 cm$^{-1}$ was converted directly to the component consisting of 810 cm$^{-1}$ as the temperature decreased.

When heating from 173 K, we can respectively observe
the increase and decrease of intensity of 825 cm$^{-1}$ and 820 cm$^{-1}$
 bands from 310K to 360K.
In DSC measurements (Fig. 1a), deviation from the baseline starts from 320K,
and reaches peak top, which indicates the end of transition, at 340K.
Since the temperature range of the transition observed in DSC and FTIR are
consistent, both measurements must detect an identical transition.
 
In addition, P3HT also showed a gradual shift of the absorption maximum from 820 cm$^{-1}$ to the lower wavenumber
(815 cm$^{-1}$) with cooling. A higher wavenumber band than 820 cm$^{-1}$ did not appear 
 as P3BT, which shows a band of 825 cm$^{-1}$ below 340 K. 
For P3DDT, a gradual shift from 820 cm$^{-1}$ to a lower wavenumber was not observed,
 but slight glowing and peak narrowing at 805 cm$^{-1}$ was observed. 
Among these polymers, only P3BT showed both a higher wavenumber shift
 around 340 K and a lower wavenumber shift between 340 K 
and 173 K with cooling.
 
A salient topic for the structure of P3ATs has been the polymorphic behavior between 
Phase I and Phase II.
However, it would be difficult to explain the IR measurement results using
only two static phases
for one reason. Although P3DDT is considered to show a transition from Phase II to Phase I around 330 K 
by the melting of alkyl chains \cite{Prosa1996,Meille1997,Causin2005}, 
the maximum absorption was maintained at 820 cm$^{-1}$ before and after the transition.
The peak shift in P3BT reflects the reasons other than the polymorphism.
  

\subsection{$^{13}$C CPMAS NMR spectra, Spin-lattice Relaxation Time, and CODEX measurements}

\begin{figure}[htbp]
\begin{center}
\includegraphics[scale=0.4]{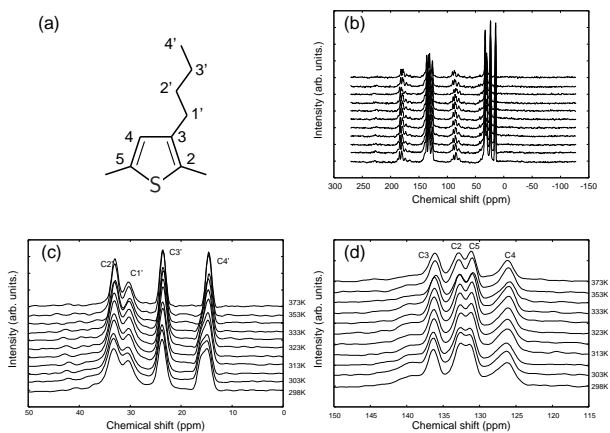}
\caption{Temperature variation of $^{13}$C CPMAS NMR spectra for P3BT. The chemical structure (a), the whole range of spectra (b), and the extended spectra for the alkyl carbons (c), and the thiophene ring (d) are shown.}
\label{fig4}
\end{center}
\end{figure}

To investigate the structure and molecular dynamics of P3BT around 340 K,
$^{13}$C CPMAS and $^{13}$C spin-lattice relaxation time measurements were performed for P3BT 
at various temperatures.
The $^{13}$C CPMAS spectra of P3BT and the chemical shift of each carbon are shown respectively in Fig. \ref{fig4}
and Table \ref{csa}.
Chemical shifts of the alkyl chain are consistent with the gauche conformation of
regiorandom poly(3-octylthiophene)(P3OT) \cite {Bolognesi2001}, indicating that the alkyl chains
behave similarly to the liquid state.
 
It is noteworthy that the clear shoulder signal of the C4'(methyl) carbon of P3BT
 is apparent around 16.0 ppm below 333 K,
 whereas the spectra at the higher temperatures show a unique component (14.6 ppm),
 meaning that at least two chemically inequivalent methyl carbons in P3BT exist below 333 K. 
Because the methylene carbons show no peak splitting,
 only the end of the butyl chain shows two or more distinct states.

In Fig. \ref{fig4}(b), the signals of the thiophene ring for P3BT
 appear over the frequency region of 120--145 ppm as previous
studies of oligothiophene \cite{Barbarella1996} and
 regiorandom poly(3-octylthiophene) \cite{Bolognesi2001} have shown. 
We observed five peaks in all measured temperature ranges (Table \ref{csa}). 
On the other hand, Bolognesi {\it et al.} \cite{Bolognesi2001} 
observed six peaks
 in the presence of both Phase I and Phase II
for regiorandom poly(3-octylthiophene).
They assigned the peaks at 140.2 ppm (C3) and 129.5 ppm
(not assigned to any carbon) to Phase II because of agreement
with the transition temperature observed by the x-ray diffraction.
In our results, although a peak is apparent at 140.2 ppm,
the peak at 129.5 ppm was not observed.
Furthermore, although the intensity of the signal decreases with heating,
the peak at 140.2 ppm exists at higher and lower temperatures than the transition temperature.
This result is controversial to Causin's argument by the x-ray diffraction
where Phase I only exist above the transition temperature.
From these results, it is difficult to explain which peaks originate
from Phase I or Phase II because two apparent
differences between
Bolognesi's and our samples are
regioregularity and alkyl chain length.
Therefore, this issue is still unclear.
 Consequently, we shall give attention only to four peaks,
 except for the peak at 140.2 ppm.
Specifically regarding the line width,
 narrowing of C4 was observed with heating.
To investigate the factors that cause this narrowing,
we carried out $T_{2 \rm H}$ measurement (see the Supplementary Information).
The spin-spin relaxation rate ($R_2$) of a rigid moiety decreases with heating,
although no sudden change was observed with respect to the transition around 340K.
Therefore, the motional narrowing of the local dipolar field occurs over the measuring
temperature range.
However, dynamic averaging of chemical shift distribution due to the twist motion
possibly causes the narrowing of $^{13}$C signal.
The latter possibility cannot be ruled out only from the $T_{2H}$ measurements.
Conclusively, we could not identify the reason for the signal narrowing
with respect to temperature elevation of the C4 carbon of thiophene.
Broadening and Splitting of C2 and C5 with heating were also observed.
We will discuss broadening later because it is expected to bear a relation with the 
spin-lattice relaxation time. 
The splitting may indicate the beginning of reorientational motion.
We will also discuss the splitting later.
Results of FTIR and CPMAS NMR show that the structural change of P3BT occurs markedly
 around 340 K.
 At temperatures greater than 340 K,
 the state of the main chain is attributed almost uniquely to 820 cm$^{-1}$ in FTIR.
 The side chain is also
 at a unique state, as shown by results of CPMAS NMR.
Below 340 K, the main chain consists of mainly two states:
 the main component attributed to 825 cm$^{-1}$ 
 and the other to 810 cm$^{-1}$ in FTIR spectra.
The methyl moiety of the side chain also shows at least two components
and is probably related to the main-chain states.   

Two questions remain. What explains the absence of the (010) peak and the weakness of 
(100) peaks below 323 K in the WAXD measurement\cite{Causin2005}?
Even if both phases (Phase I and Phase II) coexist, the respective scattering peaks of x-ray diffraction should be
observed. Furthermore, the peak shifts in the IR spectrum 
from 825 cm$^{-1}$ and 810 cm$^{-1}$ to 820 cm$^{-1}$ with heating
do not coincide with the argument by Causin {\it et al.}: the transition from Phase I and Phase II to Phase I because no common IR absorption peak attributable to Phase I is visible before and after the transition. 

The other is the driving force of the transition. 
Clarifying the driving force would be the key to organizing the information obtained from FTIR,
 CPMAS NMR, and WAXD.
Here, we specifically examine the dynamics of the twisting motion of the main chain in the crystalline state. 
Thermochromism phenomena for P3ATs are well known to result from the existence of quasi-ordered phase
or mesophase: thiophene rings have sufficient degrees of freedom to allow a twisting motion.
If the main chain had preserved the planarity without twisting until melting,
 we would not have observed gradual color changes.
The slight melting heat flow bears out the existence of the mesophase.
Additionally, the diffusion-like motion of the twisting was observed for a similar regioregular polymer,
poly(4-metylthiazole-2,5-diyl)(P4MTZ), by solid-state NMR measurements \cite{Mori2005}.
 
We performed $^{13}$C spin-lattice relaxation time measurements for P3ATs 
using Torchia method \cite{Torchia} at various temperatures to investigate the molecular dynamics.
An Arrhenius plot of the spin-lattice relaxation rate ($R_1 = T_1^{-1}$) 
for each carbon is shown in Fig. \ref{fig5}.

For the side chains (Fig. \ref{fig5}(a)), each $R_1$ value was obtained using
 a simple exponential fitting curve: 
\begin{align}
M_z(t) =  e^{-R_1t}, 
\end{align}
where $M_z$ is the magnetization along the magnetic field.
Figure \ref{fig5}(a) shows the subtle change of slopes around 333 K,
 but the decrease of $R_1$ was observed throughout the measured temperature range with heating
 (from 303 K to 373 K). 
This tendency is visible in the extremely narrowed regime 
according to the classical Bloembergen-Pound-Purcell (BPP) theory \cite{BPP},
indicating that the side chains behave similarly to liquid, which is also observed in $^{13}$C CPMAS NMR spectra.
The subtle change of slopes was inferred to result from the conformation change of the main chain
because it would be difficult to believe that further changes in average conformation should occur in a liquid state.
On the other hand, the relaxation of main chain carbons showed no single exponential decay in the measured temperature range. For that reason, we tried to fit the decay curves using a Kohlraush-Williams-Watts (KWW) function
\cite{KWW}

\begin{align}
M_z(t) \propto {\rm exp}(-\frac{t}{T_1})^\beta.
\end{align}

In general, the KWW function is used to express magnetization near and below the
glass transition temperature ($T_g$) because the distribution of the relaxation
rate gave rise to non-exponential recoveries.\cite{Schnauss1990}
Temperature dependence of $R_1$ and stretching parameter $\beta$ extracted by Eq. (2) 
for each aromatic carbons are shown respectively in Figs. \ref{fig5}(b) and \ref{fig6}. 
 Figure \ref{fig5}(b) shows that the main chain must be in the slow-motion regime because of the increase of $R_1$ with heating.
 More noteworthy is the fact that discontinuous changes are apparent around 333 K.
In many cases, a similar tendency is apparent in $^2$H NMR measurements, at the glass transition 
for many glassy formers \cite {Bohmer2001} including polymers such as polybutadiene \cite {Rossler1994}
and polystyrene \cite {Dob1998}.
In the case of NMR spin-lattice relaxation measurements,
we can distinguish three temperature regions for the arguments of glass transition. 
(i) $T < T_g$: The $R_1$ of glass formers is dominated by slow $\beta$-process (Johari-Goldstein process \cite {JG}),
 which follows the Arrhenius-type thermal activation process below $T_g$. 
(ii) $T_g < T < 1.2T_g$: Both $\alpha$ process and the Johari-Goldstein process
affect $R_1$.
Where the $\alpha$ process is described as stretched exponential functions. 
(iii) $T > 1.2T_g$: $R_1$ is dominated only by $\alpha$ process.  

Below 333 K, we observed the Arrhenius-type thermally activated process for all carbons.
That tendency is similar to the Johari-Goldstein process.
Over 333 K, although the measured temperature range is insufficient to consider the condition of (ii) or (iii) because of the limitation of available temperature range of our instrument, discontinuous changes can be a fingerprint for the effect of $\alpha$ processes.

In Fig. \ref{fig5}(b), another remarkable point exists. 
For the CPMAS $^{13}$C NMR measurements,
relaxation occurs through two paths: that by fluctuation of a local field
by $^{13}$C-$^{1}$H magnetic dipolar coupling 
and that by $^{13}$C chemical shift anisotropy (CSA).
Carbons that are directly connected to a proton, such as C4, 
have a larger dipolar coupling effect than that of unconnected carbons.
Below 333 K, only the C4 carbon shows larger $R_1$ than other carbons,
indicating that the relaxation is caused mainly 
by $^{13}$C-$^{1}$H magnetic dipolar coupling.
That is, 
the thiophene ring does not undergo reorientations sufficient to fluctuate
 the chemical shift anisotropy. 
Over 333 K, the other carbons (C2, C3, C5) showed individually specific slopes,
indicating that reorientation of the thiophene ring brings about fluctuation
 in CSA as well.
The principal components of nuclear shielding for the modeling compound
 of HT-poly(3-methylthiophene) that are
determined by GIAO-CHF also indicate the effects of CSA (Table \ref{csa_calc}).
The fact that the axis carbons (C2 and C5) with the larger $\Omega$
show steeper slopes over 333 K supports the effective CSA contribution.
Furthermore, line broadening of C2 and C5 with increasing temperature,
 as mentioned above, is explainable by the steep increase of $R_1$. That is,
the well-known nonsecular broadening \cite{Abragam}.

There are two kinds of molecular motions that could affect CSA of thiophene carbons,
translation and reorientation.
However, the possibility of translational motion would be very small,
because the diffraction peaks due to the $\pi$-$\pi$ stacking
were observed in x-ray pattern above
the transition temperature\cite{Causin2005}.
In addition, it is hard to believe that the C$_\beta$-H out-of-plane deformation mode
in FTIR
is sensitive to the translational motion.
Furthermore, the splitting of C2 and C5 signals in CPMAS NMR spectra with heating
also infers the the existence of the reorientaional motion.
Since the C2 and C5 carbons are covalently bound,
these carbons are thought of as a kind of donor-accepter pair;
namely, electronic charge densities for the C2 and C5 carbons are non-equally distributed,
resulting in opposite shift direction in $^{13}$C chemical shielding.
Taking account of both IR and NMR detectable,
the most plausible reorientation is {\it twist motion}.
The similar low energy excitations of twist motion were observed 
in structural phase transitions in oligophenylenes using Raman\cite{Guha1999} and
NMR\cite{Kohda1982,Liu1985}.
The existence of quasi-ordered phase by twist motion was also observed
in thermochromism of regiorandam P3HT\cite{Inganas}
 and the similar polymer, poly(4-methylthiazole-2,5-diyl)\cite{Mori2005}.
Since the structural modulation waves due to this twist motion show a large dispersion,
from GHz to sub kHz order\cite{Mori2005},
it would be reasonable to detectable by both NMR and IR.

The temperatures where spectral and $R_1$ slope change occur in NMR
around 323K and 333K, respectively.
These are slightly higher than DSC and IR results.
Generically the transition temperature detected by NMR is higher than
that of DSC because NMR is sensitive to the motion with higher frequency.
Therefore, NMR, DSC, and IR monitor the identical transition.

The above arguments indicate that the glass transition with respect to
the thiophene twisting in the crystalline state occurs around 333K.
We define the transition as a {\it twist-glass transition}.

However, the temperature dependence of the stretching parameter $\beta$ does not accord with the results of other glass formers.
In $^2$H NMR measurements, if the spin diffusion rate is sufficiently slower 
than the spin-lattice relaxation rate $R_1$, that is, if the recoveries of magnetization
directly reflect the molecular motions at that time,
then the experimental data are often fitted by a single stretched-exponential function like that described in Eq. (2),
from which the mean relaxation time is obtainable:

\begin{align}
\langle T_1 \rangle = \frac{T_1}{\beta} \Gamma (\frac{1}{\beta}),
\end{align}
where $\Gamma(x)$ is a Gamma function. 
For $T > T_g$, $\beta$ is nearly unity because the motional correlation time is
 much shorter than the spin-lattice relaxation time, which means that ergodicity is achieved.
Around $T_g$, the value of $\beta$ begins to decrease with decreasing temperature because of the $T_1$ distribution caused by the slowing of $\alpha$ process.
Then, the value of $\beta$ increased again slightly with further cooling
\cite{Dob1998,Hinze1996,Dob2000,Bohmer2000}. 
In these cases, the single form of the spectral density function, $J(\omega)$, is used 
to determine $T_1$ because it allows the specific examination of intramolecular motions
in ignorance of intermolecular interactions.
In our case, because of the crystalline state of strong packing, we cannot ignore
the fluctuation of intermolecular
interaction, especially above the transition temperature where twist motion exists.
For that reason, two kinds of spectral density functions are needed
 $J_{intra}(\omega)$ and $J_{inter}(\omega)$.
Because two interactions, magnetic dipolar coupling and chemical shift anisotropy, should affect each $J(\omega)$s in the specific ratio depending on the second moment 
of interactions, the decay curve cannot be expressed by a single stretched-exponential function, 
indicating that $\beta$ cannot express the $T_1$ distribution directly at higher temperatures. 
At $T < 333 K$, where a twist motion is absent,
the magnetization curve can be expressed using a single stretched-exponential function.
As mentioned above, however, at least two states exist, which are attributed to 
825 cm$^{-1}$ and 810 cm$^{-1}$ in the FTIR spectrum under the temperature region.
Therefore, the value of $\beta$ does lose the conventional 
meaning of the distribution parameter in the case of P3BT.

To detect more direct evidence of the twist-glass transition, 
we tried to perform CODEX measurements at 303K and 353K where are
lower and higher than the transition temperature, respectively.
Fig. \ref{fig7}(a) and (b) show CODEX spectra ($S$), reference spectra ($S_0$),
and pure-exchange spectra ($\Delta S = S - S_0$)
 at 303K and 353K, respectively,
with mixing time of $t_m$ = 120 ms, $t_z$ = 0.8 ms
and CSA recoupling time of $Nt_r$ = 1ms.
At 303K, no significant change is observed between $S$ and $S_0$,
consequently $\Delta S$ shows a weak signal;
that is, sufficient motion of thiophene with detectable motional amplitude during 
$ Nt_r$ (=1ms) does not occur below the twist glass transition temperature.
At 353K, on the other hand, the pure-exchange spectrum shows the signal of thiophene.
It indicates that thiophene main chain undergoes reorientation during the mixing time
($t_m$ = 120 ms) and consequently the motion attenuates recoupling of CSA.
This is a direct evidence of thiophene twisting above the twist-glass transition.
Fig. \ref{fig7}(c) shows the normalized exchange intensity as a function of 
$t_m$.
The mixing-time dependence of CODEX can be expressed as \cite{deAzevedo2000}
\begin{align}
\Delta S/S_0 (t_m) = \frac{M-1}{M} -\frac{M-1}{M}{\rm exp}(-\frac{t_m}{\tau_{\rm c}})^\beta,
\end{align}
where $\tau_{\rm c}$ is the correlation time,  $\beta$ is 
the stretching parameter,
and M is the number of equivalent orientational sites 
accessible in the motional process.
Using this equation the correlation time of thiophene twist 
is found to be $\tau_{\rm c}$ = 31.4 ms ($\tau_{\rm c}^{-1}$ = 31.8 Hz) at 353K.
However, it is difficulat to determine this value as a characteristic
 rate of thiophene twist
because the twist motion show a large dispersion 
because of the structural modulation waves as observed in the similar polymer P4MTz\cite{Mori2005}.

\subsection{Glassy crystal -- plastic crystal transition}

We were able to detect the twist-glass transition
by FTIR and $^{13}$C NMR measurements.
 The state above the transition temperature (340 K)
 should be the quasi-ordered phase with a thiophene twist.
As mentioned above,
 typical examples of the quasi-ordered phase might include liquid crystals (LCs)
 and the plastic crystals (PCs).
If the LC and PC are cooled quickly enough,
 they can be frozen respectively into a glassy liquid crystalline (GLC)
 and a glassy crystalline (GC) state.

The appearance of LC, particularly a nematic mesophase,
 in regioregular P3ATs has been reported \cite{Prosa1996,Meille1997,Causin2005}.
The phenomenon has been observed just below the melting point, where $\pi$-$\pi$ stacking
has already disappeared and only a residual order,
 caused by a side-by-side arrangement, remained.
On the other hand, below the nematic phase, P3ATs keep the $\pi$-$\pi$ stacking 
even though the alkyl side chain is in the molten state.
In this temperature range,
 the $\pi$-$\pi$ stacking is not sufficiently strong to maintain perfect planarity,
 but it is not too weak to show fluidity attributable to the LC state.  
Here, if the twist motion would be treated as a rotational freedom,
 we could define the P3ATs as a kind of PC.
Assuming P3ATs as a class of PC, we can easily interpret the twist-glass transition
 around 340 K ($\equiv T_{gp}$) for P3BT
as the transition between the glassy crystal and the plastic crystal. 

The complicated wavenumber shifts in the FTIR spectra of P3BT
are explained reasonably using this idea.
Above $T_{gp}$,
the dominant peak appears at 820 cm$^{-1}$ (Fig. \ref{fig2}a, right),
and can be assigned to the plastic crystal.
Below $T_{gp}$, two signals are observed at 825 cm$^{-1}$ and 810 cm$^{-1}$ (Fig. \ref{fig2}a, left).
The 825 cm$^{-1}$ peak is almost constant down to 173 K,
 whereas 810 cm$^{-1}$ peak increases gradually with cooling (Fig. \ref{fig3}).
We assign the 825 cm$^{-1}$ peak to the glassy crystal.
Below $T_{gp}$, the intensity of the 825 cm$^{-1}$ peak is constant, 
which is probably because the amount of GC cannot increase further 
and the twist motion of the thiophene rings is frozen.
Results of temperature-dependent WAXD by Causin \textit{et al.} \cite{Causin2005} also follow
this hypothesis: at 323 K, the (010) peak, derived from the $\pi$-$\pi$ stacking, was not observed 
because the frozen thiophene rings with various twist angles 
smear the clear x-ray diffraction. On the contrary, at 373 K, the clear diffraction peak appears because of the averaged positional order that results from the twist motion.

Next, we specifically examine the small peak at 810 cm$^{-1}$ of P3BT.
Taking account of the absence of the twist motion below $T_{gp}$, 
which is deduced from our $T_{1{\rm C}}$ measurements,
and considerably remaining a small amount of crystal from the x-ray diffraction \cite{Causin2005},
 we can assign the peaks at 810 cm$^{-1}$ for P3BT and at 815 cm$^{-1}$ for P3HT
to the crystalline phase which is constituted of the planar thiophene arrangement
with stronger $\pi$-$\pi$ stacking.
Although the reasons for a gradual transition from 820 cm$^{-1}$ to 810 cm$^{-1}$ for P3BT and to 815 cm$^{-1}$ for P3HT are not clear, 
we suggest two possible scenarios about the gradual transition.
One is that the motion of the alkyl chain is a thermally activated process 
in the extreme narrowing regime, as shown by results of $T_{1{\rm C}}$ measurements.
The process might induce the libration-like twist motion of thiophene rings. The gradual change in the IR spectra might be attributable to gradual amplification of the libration. Another explanation is the two-dimensional melting of thiophene twist via 
continuous phase transition, as expressed by
Kosterlitz-Thouless-Halperin-Nelson-Young (KTHNY) theory \cite{Kosterlitz,Halperin,Nelson,Young}.
We will discuss this problem in a future study. 

Next, we explain why the DSC curve shows not a heat capacity jump, but an endothermic peak, if the transition is a twist-glass transition.
The most reasonable explanation is that this is the effect of enthalpy relaxation of the twist-glassy state.
To examine this, we tried the DSC measurement for P3BT with various aging times
(Fig. \ref{fig8}).
The growth of endothermic peaks is observed as a function of aging time.
Typical heat capacity jump was not observed 
because the difference of heat capacity 
between glassy crystal and plastic crystal could be very small.
Another possibility of the appearance of the endothermic peak is the order-disorder transition
of thiophene ring stacking,
 which might be assigned to the peak shift from 810 cm$^{-1}$ to 820 cm$^{-1}$ in FTIR measurement.
Using that possibility, we probably cannot explain the gradual change over the wide temperature range. 
Although the transition might be too broad and its enthalpy might be too small 
to estimate the heat flow of the endothermic peak,
we conclude that the endothermic peak originates from the enthalpy relaxation
by the glassy crystal.

\subsection{Glassy crystals of P3AT with longer alkyl chains}
Why were we able to observe the twist-glass transition only for P3BT?
The most important factor is probably structural relaxation rate of the twist.
The relaxation rate is determined by several factors such as the free volume,
the mobility of both side chains and the main chain, the interaction 
between main chains, and so on.
These factors are strongly dependent on the alkyl chain length.

For P3DDT, the mobility of the main chain
would be restricted below 330 K because of the packing of the long side chains (Fig. \ref{fig1}(e)).
We infer that the twist-glass transition temperature is dependent on the side chain length,
as in the case of the conventional glass transition\cite{Cheng S.-A1992}.
From the analogy to the conventional glass transition of P3AT, 
the twist-glass transition temperature of P3DDT is presumably lower than 340 K.
Such a situation would inhibit observation of the twist-glass transition because of 
the existence of the already crystallized alkyl chains.
The FTIR spectra of P3DDT at lower temperatures support this view (Fig. \ref{fig2}(c)).

On the contrary, we expect that P3HT as well can show the twist-glass transition
because the side chain will take a liquid state over a wide temperature range, similarly to P3BT.
We observed the peak shift at 253 K from 820 cm$^{-1}$ to 815 cm$^{-1}$
with cooling in the FTIR measurements (Fig. \ref{fig2}(b), left).
 Therefore, we investigated the existence of the twist-glass transition 
that occurs around the temperature if the cooling rate is sufficiently
 high to compete with the relaxation rate of twist.
Unfortunately, the temperature controller in our FTIR instrument
is unavailable for rapid cooling and NMR is out of the available
temperature range.
Therefore, we investigated the existence of the twist-glass transition
in P3HT using DSC measurements (Fig. \ref{fig9}).
We succeeded in observing the quite small endothermic peak around 250 K
for the sample annealed for 12 h at 223 K 
after cooling from 423 K at a rate of {\it ca.} 50 K/min (Fig. \ref{fig9}(b)),
whereas the sample without annealing showed no peak (Fig. \ref{fig9}(a)).
Furthermore, the sample cooled from 533 K (over the melting temperature) at a rate of {\it ca.} 50 K/min 
shows a similar phenomenon (Fig. \ref{fig9}(c) and Fig. \ref{fig9}(d)). 
Interestingly, the temperature of the endothermic peak, as observed by the DSC measurements, 
is consistent with the temperature of the peak shift shown by FTIR measurements.
Therefore, it seems that the endothermic peak observed around 250 K
originates from the enthalpy relaxation of the glassy crystal,
as in the case of P3BT.
However, because the observed endothermic peaks are so small and 
close to the conventional glass transition (256 K) in the melt-quenched sample
by liquid nitrogen (Fig. \ref{fig9}(e)),
 it seems to be difficult to assign the transition to the twist-glass transition. 


\section{Conclusion}
Solid-state structure and dynamics of the regioregulated HT-P3ATs,
 especially P3BT,
 were investigated using FTIR and $^{13}$C NMR spectroscopies.
 The DSC measurements show that 
P3BT has an endothermic transition that occurred in the crystalline phase around 340 K. 
The FTIR measurements show that the shifts of absorption maximum for the out-of-plane mode of 
the thiophene C$_{\beta}$-H band were observed, which is consistent with DSC results.
In the $^{13}$C CPMAS NMR measurements,
 the methyl moiety shows the transition from two or more states 
to a unique state.
Furthermore, $T_{1\rm C}$ and CODEX measurements indicate that there exists
a glass transition of thiophene ring twist 
in the crystalline phase. Its enthalpy relaxation shows an endothermic peak in DSC traces. 
To P3BT,
we attempted to apply the idea that it is a glassy crystal-plastic crystal transition.
 We assigned each peak in FTIR to a specific phase. 
A phase diagram of P3BT is shown in Fig. \ref{fig10}.
At temperatures higher than the melting point ($T > \rm{540}$ K),
 the polymer is surely an isotropic liquid assigned to
836 cm$^{-1}$ in the FTIR spectrum.
 At 340 K $< T <$ 540 K,
 mostly plastic crystal (820 cm$^{-1}$) exists,
 where the polymer undergoes the twist motion of thiophene.
Below 340 K, two states exist in which the glassy crystal (825 cm$^{-1}$) has thiophene rings that are frozen with various angles 
and crystals without twisting (810 cm$^{-1}$).




\begin{centering}
\begin{table}[htmp]
\caption{Observed isotropic determined by $^{13}$C CPMAS NMR
(units in parts per million from tetramethylsilane)}
\begin{tabular}{cccccccccc}
\hline
\hline
    & \multicolumn{4}{c}{alkyl chain} & & \multicolumn{4}{c}{thiophene ring}\\
\cline{2-5} \cline{7-10}
T(K) &  $\rm C2'$ & $\rm C1'$ & $\rm C3'$ & $\rm C4'$ & &$\rm C3$ & $\rm C2$ & $\rm C5$ & $\rm C4$ \\ 
\hline
303 & 33.3 & 30.5  & 23.7  & 16.1, 14.8 & &136.4 & 132.6  & 131.4  & 126.3\\
313 & 33.2 & 30.4  & 23.6  & 15.9, 14.7 & &136.3 & 132.5  & 131.2  & 126.1\\
323 & 33.3 & 30.4 & 23.6 & 15.8, 14.6  & &136.2 & 132.7 & 131.1 & 126.1\\
333 & 33.1 & 30.3 & 23.6 & 15.7, 14.6  & &136.2 & 132.8 & 131.1 & 126.0\\
343 & 33.1 & 30.2 & 23.5 & 14.6  & &136.0 & 132.7 & 130.9 & 125.9\\
353 & 33.2 & 30.3 & 23.7 & 14.6  & &136.1 & 132.8 & 131.0 & 126.1\\
373 & 33.2 & 30.3 & 23.7 & 14.6  & &136.1 & 132.9 & 131.1 & 126.2\\
\hline
\hline
\end{tabular}
\label{csa}
\end{table}
\end{centering}

\begin{centering}
\begin{table}[htbp]
\caption{Principal components of nuclear shielding for ring carbons in HT-tri(3MeTh) calculated by the GIAO-CHF method with 6-31G(d) (units in parts per million).}
\begin{tabular}{ccccccccc}
\hline
\hline
  nucleus & $\sigma_{\rm iso}$ & $\sigma_{11}$ & $\sigma_{22}$ & $\sigma_{33}$ & $\delta$ & $\eta$ & $\Omega$ & $\kappa$\\ 
\hline
 $\rm C2$ & 73.49  & -5.40  & 73.94 & 151.94 & -78.45 & -1.01 & -157.34 & -0.009\\
 $\rm C3$ & 72.02  & -0.21  & 52.31 & 163.95 & -91.93 & -0.57 & -164.16 & 0.360\\
 $\rm C4$ & 79.72 & 6.38 & 87.31 & 145.46 & -65.74 & -1.23 & -139.08 & -0.164\\
 $\rm C5$ & 62.10 & -24.30 & 57.70 & 152.90 & -90.80 & -0.90 & -177.20 & 0.074\\
\hline
\hline
\end{tabular}
\label{csa_calc}
\end{table}
\end{centering}

\begin{figure}[htbp]
\begin{center}
\includegraphics[scale=0.4]{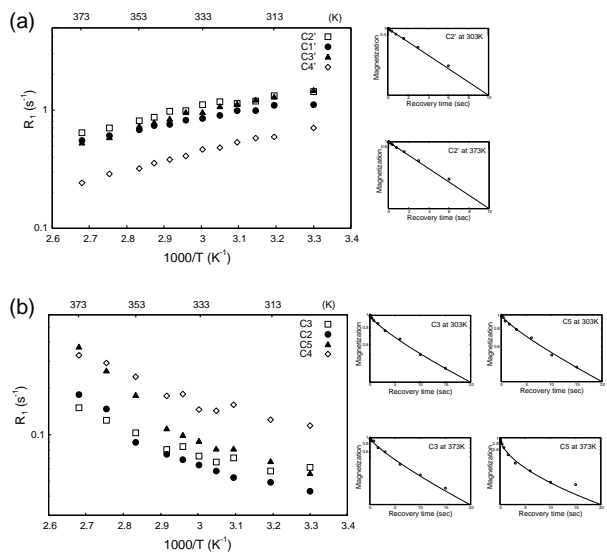}
\caption{Plots of $^{13}$C spin-lattice relaxation rate $R_1$ as a function of 1000/T for P3BT;
the alkyl side chain (a) and the main chain (b).
Typical decay curves are also shown for the butyl C2' carbon, and for the thiophene C3 and 
C5 carbons at 303 K and 373 K (insets).}
\label{fig5}
\end{center}
\end{figure}

\begin{figure}[htbp]
\begin{center}
\includegraphics[scale=0.4]{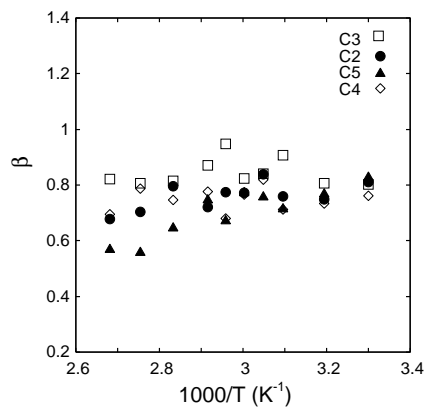}
\caption{Temperature variation of the stretching parameter $\beta$.}
\label{fig6}
\end{center}
\end{figure}

\begin{figure}[htbp]
\begin{center}
\includegraphics[scale=0.4]{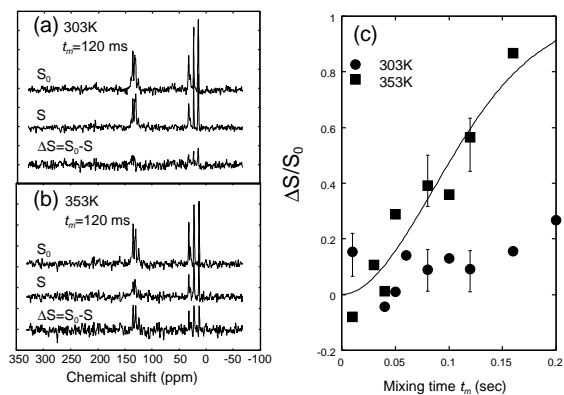}
\caption{A pure-exchange spectrum ($\Delta S$)
from a CODEX measurement (S) and a reference spectrum without exchange ($S_0$) 
at 303K (a) and 353K (b). 
$\Delta S$ was generated by subtraction of $S$ from $S_0$.
$\Delta S$/$S_0$ of thiophene twisting at 303 K and 353 K were
plotted as a function of $t_m$ (c).}
\label{fig7}
\end{center}
\end{figure}

\begin{figure}[htbp]
\begin{center}
\includegraphics[scale=0.4]{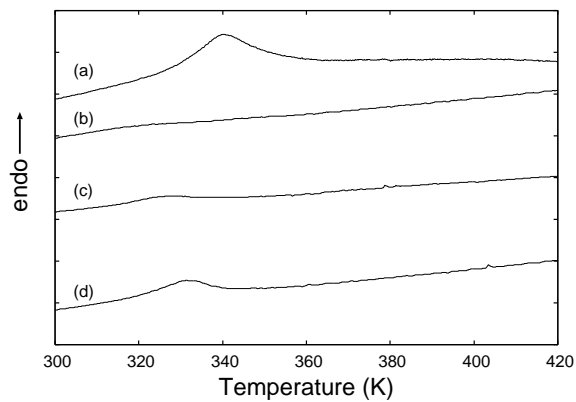}
\caption{DSC heating thermograms of the original sample (a) and 1 h storage at 298 K (b), 24 h (c), and 4 weeks (d) for P3BT after annealing at 423 K for 3 min and cooling at the rate of 10 K/min.}
\label{fig8}
\end{center}
\end{figure}

\begin{figure}[htbp]
\begin{center}
\includegraphics[scale=0.4]{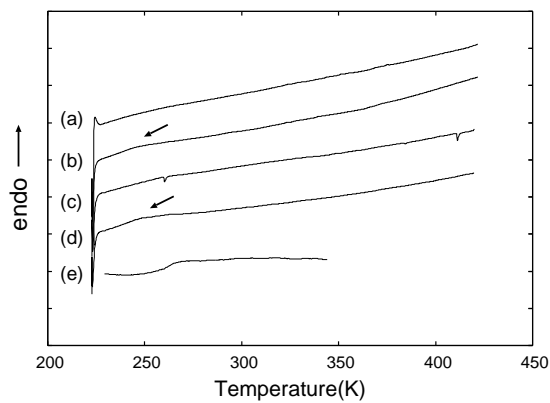}
\caption{DSC heating thermograms for the P3HT sample storaged at 
223 K for 1 min (a) and for 12 hours (b)
after cooling from 423 K at a rate of ca. 50 K/min,
 for 1 min (c) and 12 h storage (d) after cooling from 533 K 
at a rate of {\it ca.} 50 K/min, 
and for the quenched sample by liquid nitrogen from 533 K (e).}
\label{fig9}
\end{center}
\end{figure}

\begin{figure}[htbp]
\begin{center}
\includegraphics*[scale=0.4]{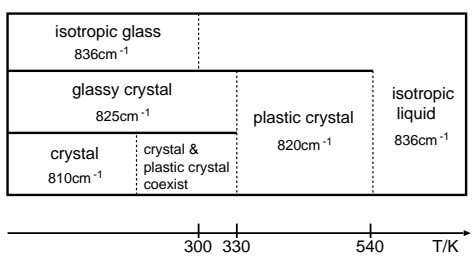}
\caption{A proposed phase diagram of P3BT. The IR absorption bands for the C$_\beta$-H bond
 out-of-plane deformation are also shown.}
\label{fig10}
\end{center}
\end{figure}

\clearpage
\section*{SUPPLEMENTARY MATERIALS}

\section{Experiments}

We carried out variable temperature proton transverse relaxation time
 ($T_{2\rm H}$) measurements,
 at 270 MHz for $^1$H 
with a home-built micro-coil probe.
The experimental conditions was set up with
90$^{\circ}$ pulse length of 2.3 $\mu$s for $T_{2\rm H}$ measurements. 
We used two-pulse Hahn echo method\cite{TPHE} for $T_{2\rm H}$ measurements.

\section{Results and discussion}

The $^1$H NMR spin echo spectra with echo time of 5$\mu$s at 303K, 333K,and  363K were shown in Fig 1(a).
We should pay attention to the broader component of spectra for investigating the C4 carbon
because the main chain must be more rigid than the alkyl side chain.
While the number of protons in the thiophene monomer unit is unity,
nine protons are in alkyl chain.
In the light of this, the broader component of the spectra originate from
not only the thiophene proton but also the root moieties of the alkyl chain
because the relative intensity of the broader component is much larger than 0.1.
Generically, the TPHE decay in rigid lattice can be fitted with

\begin{eqnarray}
M_z(t) = M(0){\rm exp}(-\frac{t}{T_2}){\rm exp}[-(\frac{t}{T_G})^2],
\end{eqnarray}
where $M(0)$,$T_2$,$T_G$, and t denote the initial transverse magnetization, the transverse relaxation time,
the characteristic decay time of the Gaussian component, and time, respectively.
In our results, however, we could not estimate the amount of the Gaussian component
because the signal from the thiophene proton was overwhelmed by signals from the alkyl protons.
There is yet another possible case where at room temperature ($T<R_{gp}$) 
the thiophene proton is decoupled with
the surrounding protons due to rapid molecular motion of alkyl chains.
In this case, the signal of the thiophene proton should have a Lorentzian shape.
In any cases, the decay of broader signal must be originated from multiple components
because of signals derived from both thiophene and alkyl protons.
Therefore, we fitted the decay with a sum of two Lorentzian curves, given by
\begin{eqnarray}
M_z(t) = M_f(0){\rm exp}(-\frac{t}{T_{2f}})+M_s(0){\rm exp}(-\frac{t}{T_{2s}}),
\end{eqnarray}

where $T_{2f}$ and $T_{2s}$ denote lifetimes for faster and slower decays, and $M_f(0)$ and
$M_s(0)$ are the respective intensity at t=0.
Since the proton in thiophene must be included in faster decay component,
we carried out the Arrhenius plot of spin-spin relaxation rate ($R_2 = T_2^{-1}$)
of the broader signal (in Fig.1(b), typical decay curves at 303K and 363K are also shown
in insets).
The value of $R_2$ decreases with heating,
although no sudden change was observed with respect to the transition around 340K.
Therefore, the motional narrowing of the local dipolar field occurs over the measuring
temperature range.
However, in $^{13}$C CPMAS spectra, dynamic averaging of chemical shift distribution due to the twist motion
possibly causes the narrowing of the C4 signal.
The latter possibility cannot be ruled out only from the $T_{2H}$ measurements.
Conclusively, we could not identify the reason for the signal narrowing
of the C4 carbon of thiophene in association with temperature elevation.

\begin{figure}[htbp]
\begin{center}
\includegraphics[scale=0.3]{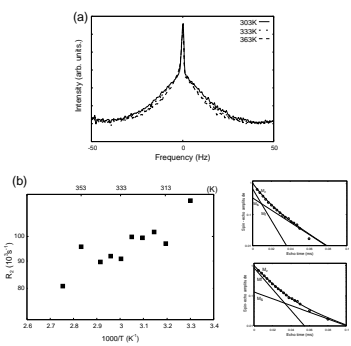}
\caption{$^1$H two-pulse Hahn echo spectra with echo time of 5$\mu$s
at 303K, 333K, and 363K (a).
Plots of $^{1}$H spin-spin relaxation rate $R_2$ of the broader component 
as a function of inverse temperature, 1000/T (b).
 Typical decay curves at 303K and 363K are also shown
in insets.
All decay curves were normalized so that the value at zero time equals unity.
}
\label{fig5}
\end{center}
\end{figure}

\end{document}